# The Ligo-Virgo-KAGRA Computing Infrastructure for Gravitational-wave Research

*Stefano* Bagnasco[1*] for the Virgo Collaboration and the LIGO Scientific Collaboration

[1]INFN Sezione di Torino, 10125 Torino, Italy

**Abstract.** The LIGO, VIRGO and KAGRA Gravitational-wave (GW) observatories are getting ready for their fourth observational period, O4, scheduled to begin in March 2023, with improved sensitivities and thus higher event rates. GW-related computing has both large commonalities with HEP computing, particularly in the domain of offline data processing and analysis, and important differences, for example in the fact that the amount of raw data doesn't grow much with the instrument sensitivity, or the need to timely generate and distribute "event candidate alerts" to EM and neutrino observatories, thus making gravitational multi-messenger astronomy possible. Data from the interferometers are exchanged between collaborations both for low-latency and offline processing; in recent years, the three collaborations designed and built a common distributed computing infrastructure to prepare for a growing computing demand, and to reduce the maintenance burden of legacy custom-made tools, by increasingly adopting tools and architectures originally developed in the context of HEP computing. So, for example, HTCondor is used for workflow management, Rucio for many data management needs, CVMFS for code and data distribution, and more. We will present GW computing use cases and report about the architecture of the computing infrastructure as will be used during O4, as well as some planned upgrades for the subsequent observing run O5.

## 1 Introduction

The very successful third observational period (O3, from April 2019 to March 2020) of the LIGO [1] and Virgo [2] gravitational-wave observatories collaboration led to many first observations, and the publication of three "catalog" papers [3,4,5] that complement the one including events from previous runs O1 and O2 [6], shifting the focus from discovery to actual study of the extreme events that generate the waves. Among the 90 high-significance events collated in the papers there are the first observation of an asymmetrical-mass binary black hole (BBH) merger [7], two mergers of a black hole with a neutron star [8], a merger of two very large black holes with a total mass of 150 M$_\odot$ [9], and more.

Between observing runs, the collaborations have upgraded their detectors to reach higher sensitivities. The sensitivity is commonly expressed as "range", i.e., the distance at which a binary neutron star merger can be detected. So, for example, the LIGO range was 80 Mpc for O1, 100Mpc for O2 and 100-140 Mpc for O3. Since the volume of space observed grows with the third power of the range, the rate of expected events grows likewise. A characteristic of gravitational-wave data is that the size of the data used for final analyses does not grow with the sensitivity, and even though the size of the raw data does increase with the complexity of the instrument, data management does not become a huge problem with upgraded detectors. However, the amount of computing power needed to extract the events grows, and the problem is exacerbated by the need to analyse them promptly to produce public alerts for multimessenger observation.

---

[*] Corresponding author: stefano.bagnasco@to.infn.it

The LIGO, Virgo and KAGRA [10] collaborations are now getting ready to start their fourth observational period O4, scheduled to start in March 2023, with upgraded detectors and an even higher sensitivity[†].

## 1.1 Low-latency and Offline computing domains

One of the main differences between gravitational-wave data processing and, for example, High Energy Physics computing is the existence of *three* distinct computing domains.

The first one is "online": data acquisition and preparation, the aggregation in manageable chunks and the production of reduced sets of data for different uses happen in quasi-real time, in dedicated computing infrastructures close to the detectors. While there is no concept of "trigger" in interferometric data acquisition, in many aspects this is not conceptually different from what is done in other fields of experimental physics and is anyway outside the scope of this paper.

At the other end of the spectrum there is asynchronous "offline" analysis: pre-processed data, stored off-site often in shared computing centres, are usually analysed using conventional batch processing, with workloads being executed, possibly (as we will see) on a distributed infrastructure. The typical payload mostly consists of deep searches for several categories of signals, detector characterization activities and some simulation.

The third computing domain is "low-latency", which is actually common to the whole time-domain astronomical community. One of the most promising and interesting fields today is multimessenger astronomy, i.e., the possibility to observe an astrophysical event using more than one messenger: electromagnetic radiation in different energy bands, neutrinos and, indeed, gravitational waves. Doing so requires the timely exchange of alerts between observatories, to promptly point the instruments, make sure the data is recorded for subsequent analysis or take whatever action is required to observe the event.

Since most gravitational-wave analyses are only meaningful if performed on data from more than one detector (e.g., when coherence methods are applied for the detection of unmodelled burst signals), and in any case coincidence between detectors is needed to reduce effects of noise and triangulate the sky position of the source, LIGO, Virgo and KAGRA agreed to exchange their data also in low-latency and search for event candidates together. Thus, when in "observing mode" (i.e., with stable detectors generating useful credible data) data are transferred between the centres where they are analysed, on dedicated computing clusters, to generate alerts. Whenever a candidate event is detected, its summary data are stored in a publicly accessible dedicated database and an alert is distributed, with updates as more precise sky location and parameter estimation are performed; this is described in more detail in Sec. 3.

For the last few years, the three collaborations joined their computing efforts and built a common infrastructure, serving the International Gravitational-Wave observatory Network (IGWN), currently formed by the LIGO, Virgo and KAGRA interferometers. As described in [11], the increasing computing requirements of upgraded detectors with a higher event rate and the need to reduce the maintenance effort on custom tools prompted a shift towards widely used, mainstream tools and the use of shared, distributed computing resources. This is described in Sec. 4.

---

[†] Due to delays in the commissioning of the interferometers, O4 started on May 24[th], 2023, with only the two LIGO detectors active. At the time of writing, Virgo and KAGRA are planned to join at a fac

## 2 Data management and distribution infrastructure

Gravitational-wave data are relatively simple in structure. Raw data from each interferometer is composed by time series of data from o($10^5$) channels (detector components, detector monitoring instruments, seismometers, magnetometers, etc.), from which calibrated data are produced, including the single physics channel, the dimensionless "strain" or $h(t)$ and synthetic state information. All these make up a flux of o(50 MB/s), or about 120 TB per month of observation (numbers differs slightly across the network, since the instruments are different).

So-called "aggregated $h(t)$" files used for the low-latency and offline analyses (with different file sizes) are then produced, in the proprietary GW Frame File format [12], along with a few types of downsampled and summary data for detection characterization studies or other uses. The total size of the data used for final offline analyses is about 10TB per year of observation from each interferometer. Data quality metadata are stored in the DQSEGDB [13], that also provides an HHTP API to query and retrieve information about the status of the detectors at any relevant time interval.

After a proprietary period, aggregated $h(t)$ files are published on the Gravitational-Wave Open Science Centre [14,15] for public use.

### 2.1 Low latency

As already mentioned, data are exchanged in quasi-real time between the collaborations to perform low-latency data analyses on dedicated clusters (see sec. 3). To reliably manage the transfers, a software tool has been developed in the context of the IGWN, known as IGWN Low-Latency Data Distribution (igwn-lldd) [16]. The *igwn-lldd* library leverages the power of Apache Kafka [17] to facilitate seamless data transfer across the sites, ensuring convenient access to the data stream, reliability, scalability, and very easy reconfiguration of the network topology.

Currently data are transferred from the active detector sites (LIGO Hanford, LIGO Livingston, Virgo, KAGRA) to "production" processing clusters at the LIGO Laboratory at Caltech (comprising about 15k CPU cores and 400 GPUs) and at EGO (about 1k CPU cores), and number of other computing centres, including Penn State University and University of Wisconsin at Madison in the US, INFN-CNAF in Italy and CIEMAT in Spain.

### 2.2 Offline

Data for offline processing are distributed to external Computing Centres through the Open Science Data Federation [18] (OSDF), a content delivery network for scientific data developed and maintained by the OSG Consortium [19] and based on the xrootd data management framework [20]. In that architecture, data are published by transferring them to "Origin" servers and distributed through a series of cache instances.

The data namespace is distributed to centres via CVMFS [21], and can be discovered via GWDataFind [22], a service that scans repositories of data in the GW Frame File format and return lists of file access URLs according to criteria defined in a query by the client. Data are then physically accessed by reading them from the closest OSDF cache instance either through CVMFS's *externalData* feature or directly through the xrootd protocol using the OSDF client tools, such as the OSDF client [23] or the HTCondor file transfer plugin.

Currently the network, depicted if Fig. 1, comprises five Origin servers: one for each interferometer proprietary data (the one for LIGO is hosted at the LIGO Laboratory at Caltech, the one for Virgo data at Université Catholique de Louvain in Belgium, while KAGRA is in the process of setting one up); one for the Open Data managed by the LIGO

Laboratory at Caltech for the GWOSC, and two (one at Caltech and one at the SurfSARA computing centre in Amsterdam, mirroring each other) for pre-processed data (e.g., Fourier-transformed $h(t)$ for analyses in the frequency domain) used by some pipelines.

Data transfers to the Origin servers are principally managed using the Rucio data management framework [24], even though some long-standing transfers (e.g., Virgo raw data transfer from EGO to custodial storage at CNAF and CC-IN2P3) are still managed with proprietary tools.

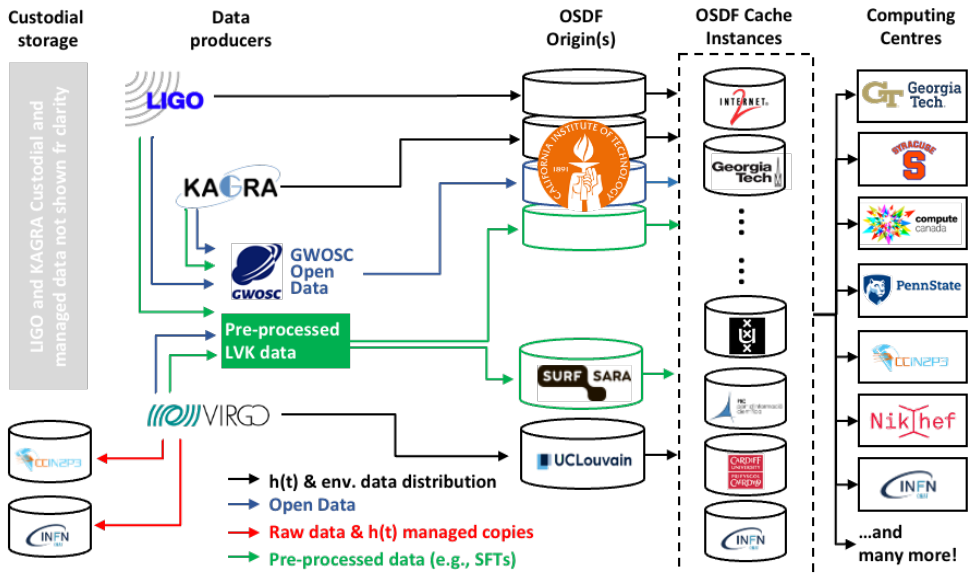

**Fig. 1**. Higher-latency data distribution infrastructure.

## 3 Low-latency infrastructure and services

Several low-latency search pipelines run on dedicated clusters at CIT and EGO, looking for transient gravitational-wave signals. Event candidates are fed to the IGWN Low Latency Alert Infrastructure (LLAI), that handles the generation and distribution of astronomical alerts.

The main components of the LLAI are:

- the GWCelery [25] service that manages the computational processes that produce the information included in the alerts;
- the GraceDB [26] events database;
- the SCiMMA Hopskotch [27] messaging system for communication between the components and actual alert distribution.

Candidate events are uploaded by search pipelines to GraceDB, which in turn posts a message in a dedicated Hopskotch topic. Upon reception of the message, GWCelery schedules rapid analyses of the data and uploads the results back to GraceDB. Every candidate event is thus "enriched" by adding information obtainable from computationally cheap tasks, such as checks on data quality or detector status. Before performing more computationally intensive tasks, separate candidates that are likely coming from the same astrophysical source are grouped together in superevents. For each superevent, such computationally intensive tasks like sky localization or parameter estimation are performed, and the superevent is enriched with the new details.

GWCelery then sends out the public alerts as a machine-readable public notice in JSON and Apache Avro [28] formats via the Hopskotch server (on a separate topic) and via the NASA General Coordinates Network (GCN) [29], in the form of human-readable plain text messages. Alerts are also produced in the VOEvent XML format [30] developed by the International Virtual Observatory Alliance (IVOA) and distributed through VOEvent brokers. Information and data about the event candidates are available via the GraceDB public interface, shown in Fig. 2.

**Fig. 2**. The public GraceDB web interface.

Preliminary alerts are issued through SCiMMA only, usually within ~1 min after the candidate event is detected [31], and the "initial" alert (or a retraction) is issued after human vetting of the candidate. Alerts are also sent out with updated information when available. In some cases (particularly loud and close-by binary neutron star mergers, with signals staying for minutes in the sensitivity band of the instruments), it should even be possible to issue "early warning" alerts up to a minute *before* the merger.

## 4 Offline workflow management system

Historically many gravitational-wave analysis pipelines were designed to run in a specific environment, such as the LIGO cluster at CalTech or the CC-IN2P3 computing centre in Lyon. A big effort has been spent in recent years to migrate to a more flexible use of computing resources, including exploitation of opportunistic resources. The first step toward this was the design and implementation of a common architecture, which is composed by the data access interfaces described above and a common workload management infrastructure based on the widely used HTCondor [32] software. A significant fraction of the offline analysis is still done via local submission to private HTCondor pools, but the recommendation to analysis groups is to use the IGWN Pool: a large, distributed pool made up of dedicated computing resources provided by IGWN member groups as well as opportunistic resources from external partners in the US, Europe and elsewhere, connected through support infrastructure maintained in partnership with the Open Science Grid. The contribution provided by each cluster in the IGWN Pool for the processing of O3 data is shown in Fig. 3; the total amount of CPU used for O3 data processing was more than $7.5\times10^9$ HS06 hourswith demand expected to grow by at least 50% for O4.

Fig. 4 shows the overall breakdown of resource consumption, while Fig. 5 shows a timeline of resource usage by observing run.

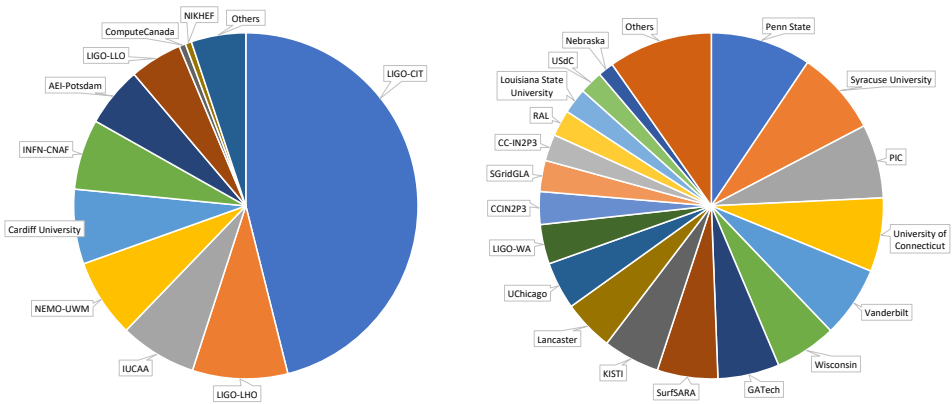

**Fig. 4**. Left: relative contributions of computing centres to the processing of O3 data. Right: disaggregation of the "Others" wedge from the left pie chart. Overall, more than 50 centres contributed to the effort.

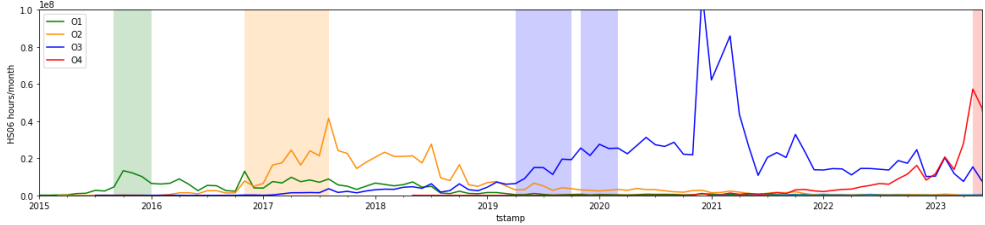

**Fig. 5**. Overall CPU resource usage timeline, by observation period tag. Shaded areas indicate the duration of the observing runs, vertical axis units are HepSPEC06 hours per month.

Code is distributed through CVMFS either in the form of Conda environments [33], that are manged through conda-forge [34], or AppTainer (formerly known as Singularity) [35] containers.

Submission of the workflows to the IGWN Pool, either as single jobs or DAGs, is via dedicated HTCondor Access Points. For accounting and prioritization purposes, each job is labelled with a tag (using HTCondor's *accounting_group* feature) describing the type of work being performed, e.g.:

```
ligo.prod.o3.cbc.imbh.gstlaloffline
```

describes a "production" (i.e., not development, test or simulation) job analysing data from the O3 observing run, by the Compact Binary Coalescence (CBC) group; the search pipeline is one based on the GstLAL [36] package looking for signals from intermediate mass black holes (IMBH). Job accounting data are queried daily from Access Points and aggregated in a database.

## 5 Conclusions

The LIGO, Virgo and KAGRA collaborations (as the LVK Collaboration) have designed and are building and maintaining the International Gravitational-Wave observatories Network computing infrastructure, a distributed computing infrastructure catering to the low-latency and offline computing needs of the collaborations.

The infrastructure has been designed by exploiting as much as possible standard and widely used tools, and minimizing the amount of custom software, to reduce the maintenance burden and be able to integrate in the wider physics computing community.

Even though much of the computing workload is still being run locally on dedicated clusters, more and more analyses are carried out on the distributed infrastructure, while more resources are added to the pool. An earlier version of the infrastructure successfully provided the resources for the processing of O3 data, and the updated version is being used for the upcoming O4 run.